\documentclass[a4paper,11pt]{article}

\usepackage{pos}  
\usepackage{sidecap}
\usepackage{amsmath}
\usepackage{wrapfig}
\usepackage{gensymb} 
\usepackage{hyperref}
\usepackage[capitalize]{cleveref}
\usepackage{url} 
\newcommand{\fixme}[1]{}

\newcommand{\Sonetwentyfive}{$S_{125}$}

\newcommand{\sectionvspacecheat}{\vspace{-0.5em}}

\title{Advances in reconstructing the cosmic-ray energy spectrum with IceTop}
\ShortTitle{IceTop Advances in Reconstruction}

\author{The IceCube Collaboration \\{\normalsize \normalfont(a complete list of authors can be found at the end of the proceedings)}\\}

\emailAdd{lilly.pyras@utah.edu}
\emailAdd{larissa.paul@sdsmt.edu}
\emailAdd{krawlins@alaska.edu}
\emailAdd{ian.reistroffer@mines.sdsmt.edu}
\emailAdd{amar.thakuri@mines.sdsmt.edu}

\abstract{
The IceTop array at the surface of the IceCube Neutrino Observatory measures 
extensive air showers produced by cosmic-ray particles with energies from PeV up to EeV, covering the transition region from galactic to extragalactic sources. This contribution presents significant improvements that will enhance the measurement of the cosmic-ray energy spectrum. 
(I) To analyze more than a decade of data with increasing snow overburdens on the detector, an improved method to handle the time-dependent attenuation of the detector signals was developed. 
(II) New analysis cuts have been developed to increase the measured event rate while improving the reconstruction quality. 
(III)~A~new reconstruction method that separately fits the electromagnetic and muonic components, while incorporating data from the deep in-ice detector, enables the reconstruction of air showers with cores outside the IceTop array, up to zenith angles of 50\degree.
These improvements will significantly increase the event statistics and extend the IceTop spectrum towards higher energies.

\vspace{4mm}

{\bfseries Corresponding authors:}
Lilly Pyras$^{1*}$, 
Larissa Paul$^{2}$, 
Katherine Rawlins$^{3}$, 
Ian Reistroffer$^{2}$, 
Amar Thakuri$^{2}$\\
{$^{1}$ \itshape Department of Physics and Astronomy, University of Utah}\\
{$^{2}$ \itshape Physics Department, South Dakota School of Mines and Technology}\\
{$^{3}$ \itshape Department of Physics and Astronomy, University of Alaska Anchorage}\\[4mm]
$^*$ Presenter
}

\ConferenceLogo{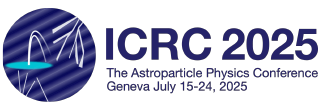}

\FullConference{39th International Cosmic Ray Conference (ICRC2025)\\
 15–24 July 2025\\
Geneva, Switzerland\\}

\begin{document}

\maketitle

\sidecaptionvpos{figure}{c} 

\section{Introduction and Detector}
\label{sec:1_intro} 
\sectionvspacecheat
When high-energy cosmic rays enter Earth's atmosphere, they 
produce cascades of secondary particles known as extensive air showers (EASs). IceTop, the surface array of the IceCube Neutrino Observatory, detects these air showers in the PeV to EeV energy range, 
covering what is
thought to be the transition region from galactic to extragalactic in origin~\cite{PhysRevD.110.030001}.
This work addresses some of the statistical and systematic limitations of previous spectrum analyses~\cite{3year_composition_paper, IceCube:2020yct}, especially at the highest energies, by introducing a data-driven snow model, improved event selection, and a new reconstruction for events with a shower core outside of IceTop. These methods lay the groundwork for a large variety of future analyses in IceTop.

The IceTop array \cite{IceCube:2012nn} is comprised of 81 ``stations'', each containing two tanks (A and B) equipped with two Digital Optical Modules (DOMs). 
EASs are detected through the Cherenkov light produced by charged particles passing through the ice inside the tanks. 
If both tanks in a station register a signal, the hits are recorded as Hard Local Coincidence (HLC);
a hit is Soft Local Coincidence (SLC) if only one tank recorded a signal.
Signals are calibrated to units of vertical equivalent muons (VEM).
Reconstruction of EAS events employs a maximum-likelihood method.
The expected charge $S$ at a distance $r$ from the shower axis is modeled with a lateral distribution function (LDF), in particular: a `Double Logarithmic Parabola' (DLP) \cite{DLPRef}:
\begin{equation} \label{eq:Sem}
S(r) = S_{\text{ref}} \cdot \left( \frac{r}{r_{\text{ref}}} \right)^{-\beta - \kappa \cdot \log_{10} \left( \frac{r}{r_{\text{ref}}} \right)}, \ r_{\text{ref}} = 125\,\rm{m}.
\end{equation}
The expected arrival times of the signals---the geometric shape of the shower front---is modeled by a paraboloid with a Gaussian nose \cite{IceCube:2012nn}.
From these models, the core position and direction of the shower can be reconstructed, as well as the ``shower size'': the signal $S_\mathrm{ref}$ measured at the reference distance $r_\mathrm{ref}$ from the shower axis. 

Previous analyses used only HLC hits for reconstruction, and chose a reference distance of 125~meters, with \Sonetwentyfive{} used as a proxy for the primary energy. 
The reference distance was motivated by earlier studies with a smaller detector~\cite{kislat2011}, dividing IceTop into Subarrays A and B using one tank per station. Recently, this technique has been extended to the full-size detector and higher energies, to study alternative reference distances for air showers with cores located both inside the IceTop array (contained events) and outside of it (uncontained events).
By fitting $S_{\text{ref}}$ and $\beta$ separately for Subarray-A and B across different $r_{\text{ref}}$ values, we identify the distance that minimizes fluctuations between subarrays.
The initial results indicate that the best $r_{\text{ref}}$ at all energies is higher for uncontained events compared to contained events. Additionally, for both event types, the optimal $r_{\text{ref}}$ increases with energy. This suggests that the most stable $S_{\text{ref}}$ lies near the distance where the bulk of the signal information is concentrated relative to the shower core. These findings may offer a more precise approach for LDF reconstruction and subsequently resolution, particularly at higher energies.
\section{Snow}
\label{sec:2_snow}
\begin{figure}[t!]
\centering
\includegraphics[width=0.4\linewidth]{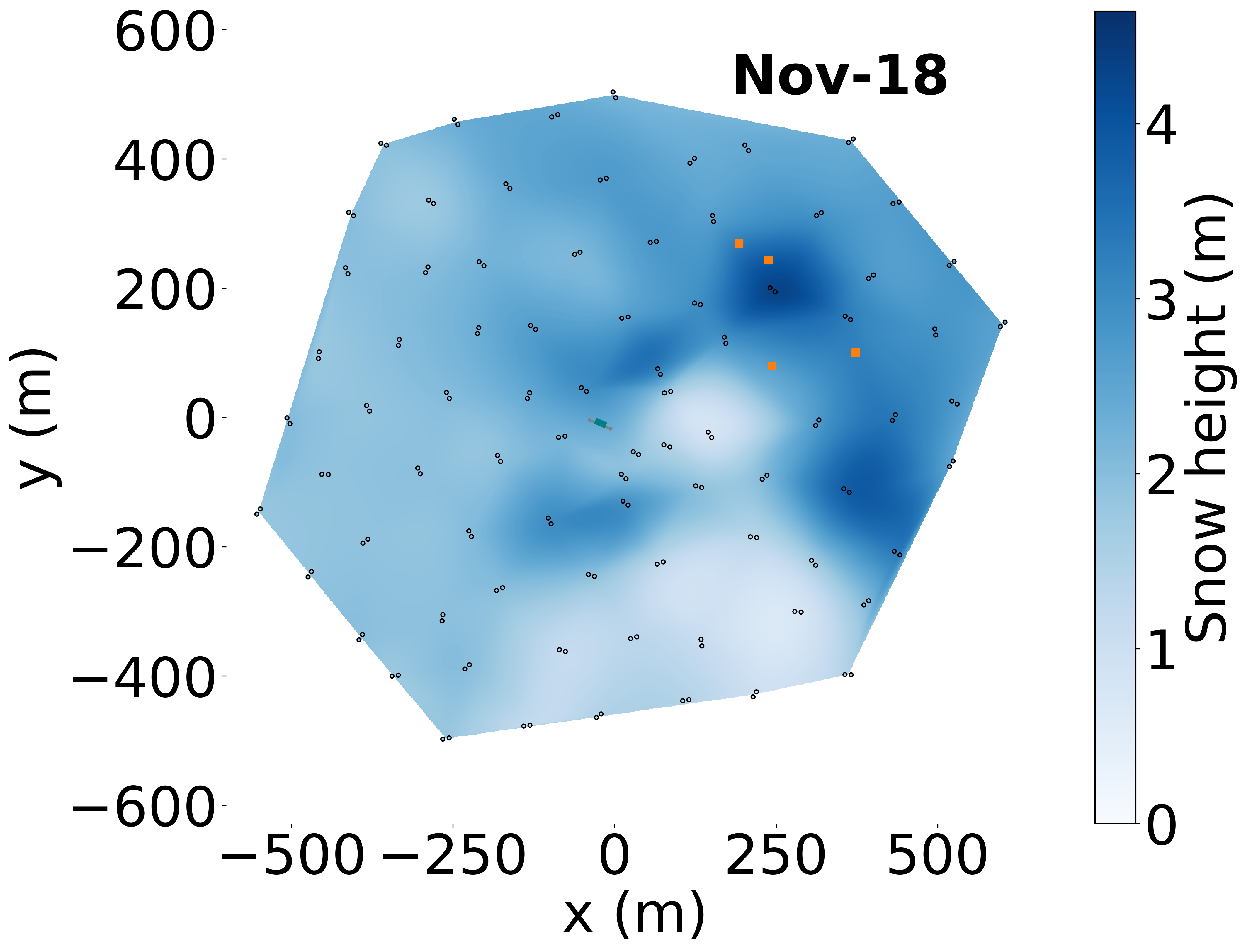}
\hspace{0.05\linewidth}
\includegraphics[width=0.4\linewidth]{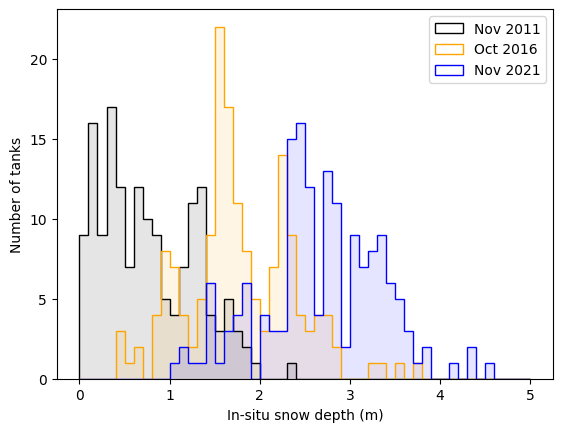}
\caption{In-situ snow depths of the tanks in IceTop: 
a) over the array at one particular time, and 
b) sampled over three disparate times.
\fixme{Do we need the left one? To-do: Change "Nov-18" to "Nov-2018".}
}\label{fig:snow_over_time}
\end{figure}

\sectionvspacecheat
The IceTop tanks were deployed in stages over several years ending in 2011, with each tank initially installed flush with the snow surface. Since then, wind-blown snow has steadily accumulated across the array, creating an uneven and time-dependent overburden that varies from tank to tank, as illustrated in \cref{fig:snow_over_time}. This non-uniform snow cover alters the detector response and makes the reconstruction of air showers inconsistent across different years of data. 
Attenuation is expected to depend on the energies of the shower particles, which may itself depend on the shower size and the distance
from the shower's core. Thus,
to achieve a uniform reconstruction across multiple years, a snow attenuation model must 
be applicable in a wide range of shower conditions.
The previous model, 
``RADE--3''~\cite{Rawlins:icrc2023},
is used as a basis for its successor, ``RADE--4''.

\begin{figure}[t!]
\centering
\includegraphics[width=0.33\linewidth]{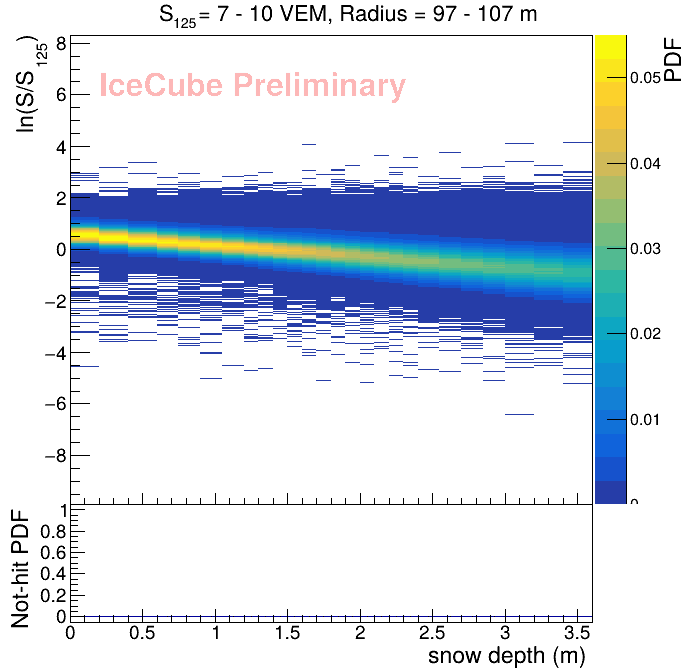}
\includegraphics[width=0.33\linewidth]{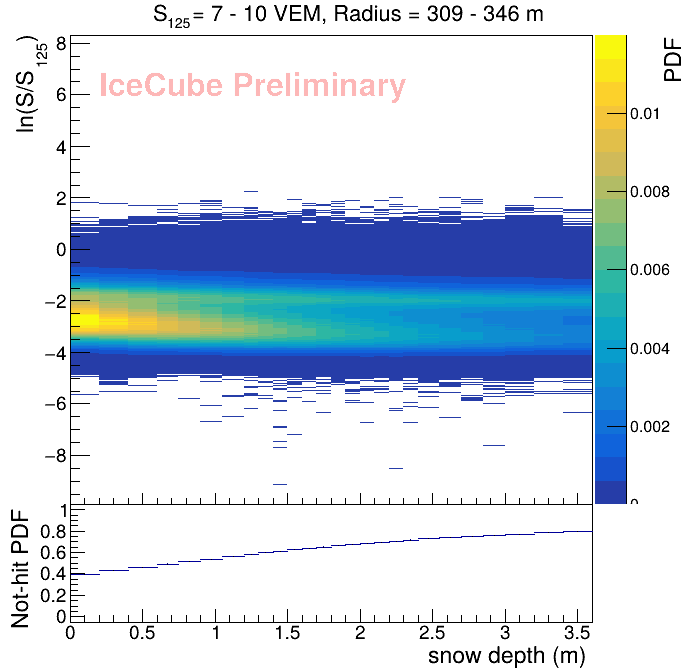}
\includegraphics[width=0.28\linewidth]{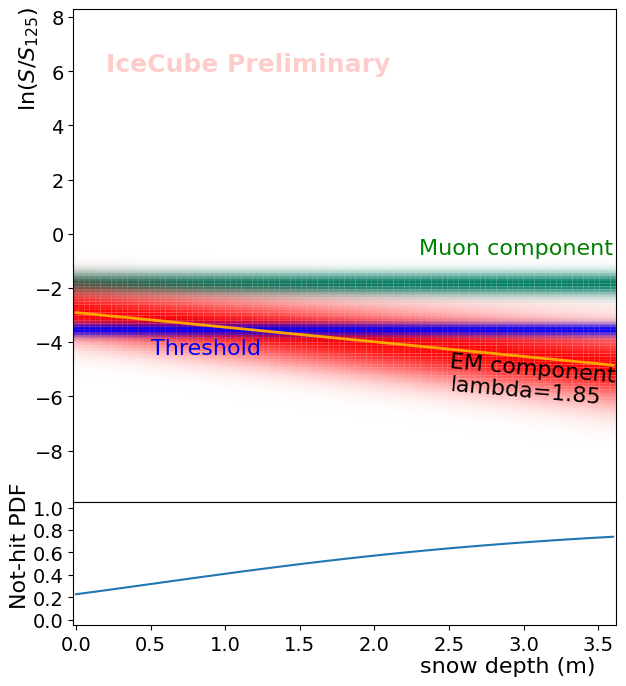}
\caption{Two-dimensional histograms of log-normalized charge vs.~snow depth, for two different distances
from the core.
Left: at $\sim$100 meters, the attenuation of the snow can be easily measured. 
Center: at $\sim$290 meters, the electromagnetic component drops below the detector threshold, and the non-attenuating ``muon peak'' is also visible. 
Right: A visualization of the 2-D model fitted to the charges in the center figure, with an EM component (red) and a muon component (green). The effect of a detector threshold (blue) populates the not-hit ($S=0$) component of the distribution (the lower panels).
}
\label{fig:snow_2dmap}
\end{figure}

In this data-driven method, IceTop tank charges from 11 years of data (2011 through 2021) 
are collected and binned as a function of multiple parameters:
the shower size (\Sonetwentyfive), the perpendicular distance of the tank from the shower core ($r$), and the slant depth through the snow to the tank ($d = z_{\text{snow}}/\cos(\theta)$). Each tank's charge ($S$) is normalized by the \Sonetwentyfive{} of the shower it appears in, to make a ``log-normalized charge'': $S_{\ln} = \ln(S/S_{125})$. Not-hit tanks ($S=0$) are also recorded, 
in a specially-designated bin in log-normalized charge.
Within each bin of \Sonetwentyfive{} and $r$, the log-normalized charges are mapped as a function of snow slant depth. \cref{fig:snow_2dmap}~(left) shows a straightforward example of such a map; 
$\ln(S/S_{125})$
decreases at a rate that corresponds to a characteristic attenuation length $\lambda$.
At large radii (\cref{fig:snow_2dmap}, center), 
the situation gets more complicated in two ways: a) the electromagnetic (EM) charges drop below the tank's threshold and are recorded as non-hits ($S=0$) instead of as hits, and b) a second structure appears in the charge distribution, from muons. The ``muon peak'' is largely unaffected by snow attenuation. 
%
%
In this work, a series of 2-D histograms---each such as \cref{fig:snow_2dmap}'s left panels---are fit to a model in which there is both an EM component (with fraction $F_\text{EM}$) and a muon component (with fraction $1-F_\text{EM}$). Both are modeled as Gaussians in log-normalized charge $G(S_{\ln},\mu,\sigma)$. The mean and sigma of the non-attenuating muon component ($\mu_\mu$ and $\sigma_\mu$) are constant with snow depth. But $\mu_\text{EM}$ is modeled to fall off with a characteristic attenuation length $\lambda$, and $\sigma_\text{EM}$ is modeled to vary linearly with some slope. The detector threshold is modeled as a sigmoid function $Y$ with a central position $S_\text{thresh}$ and a width of $\delta$. So the function fit is:
\vspace{-1em}
\begin{equation}\label{eq:full_model}
\begin{split}
      f\left(S_{\ln},d\right)= (1-F_{\text{EM}})G(S_{\ln},\mu_{\mu},\sigma_{\mu}) + F_{\text{EM}}Y(S_{\ln},S_{\text{thresh}},\delta) G(S_{\ln},\mu_{\text{EM}},\sigma_{\text{EM}}) \\
    ~~~~~\text{where:}~~~~~
    \mu_{{\text{EM}}} = \mu_{\text{EM}_{\text{no snow}}}- d/\lambda,
    ~~~~~\text{and}~~~~~
    \sigma_{\text{EM}} = \sigma_{\text{EM}_{\text{no snow}}}+\sigma_{\text{EM}_{\text{slope}}}d.
\end{split}
\end{equation}
The number of non-hits at $S=0$, which is recorded in its own designated bins of the histogram, is modeled as the number of hits expected to fall below the threshold:
\begin{equation}\label{eq:EM_zeros}
    f_{\text{zeros}} =
    \frac{1}{2}\left[ 1+\text{erf}\left( \frac{S_{\text{thresh}}-\mu_{{\text{EM}}}}{\sigma_{\text{EM}}\sqrt{ 2 }} \right) \right].
\end{equation}
The hits and non-hits for the entire 2-D histogram are fitted together. Including the non-hits in the fitting helps greatly to constrain the relationship between the EM Gaussian and the detector threshold. \cref{fig:snow_2dmap}~(right) shows a visualization of the various components of this fit (muon, EM, and threshold), including the non-hits in a separate panel.

\begin{SCfigure}[][b]
\includegraphics[width=0.5\linewidth]{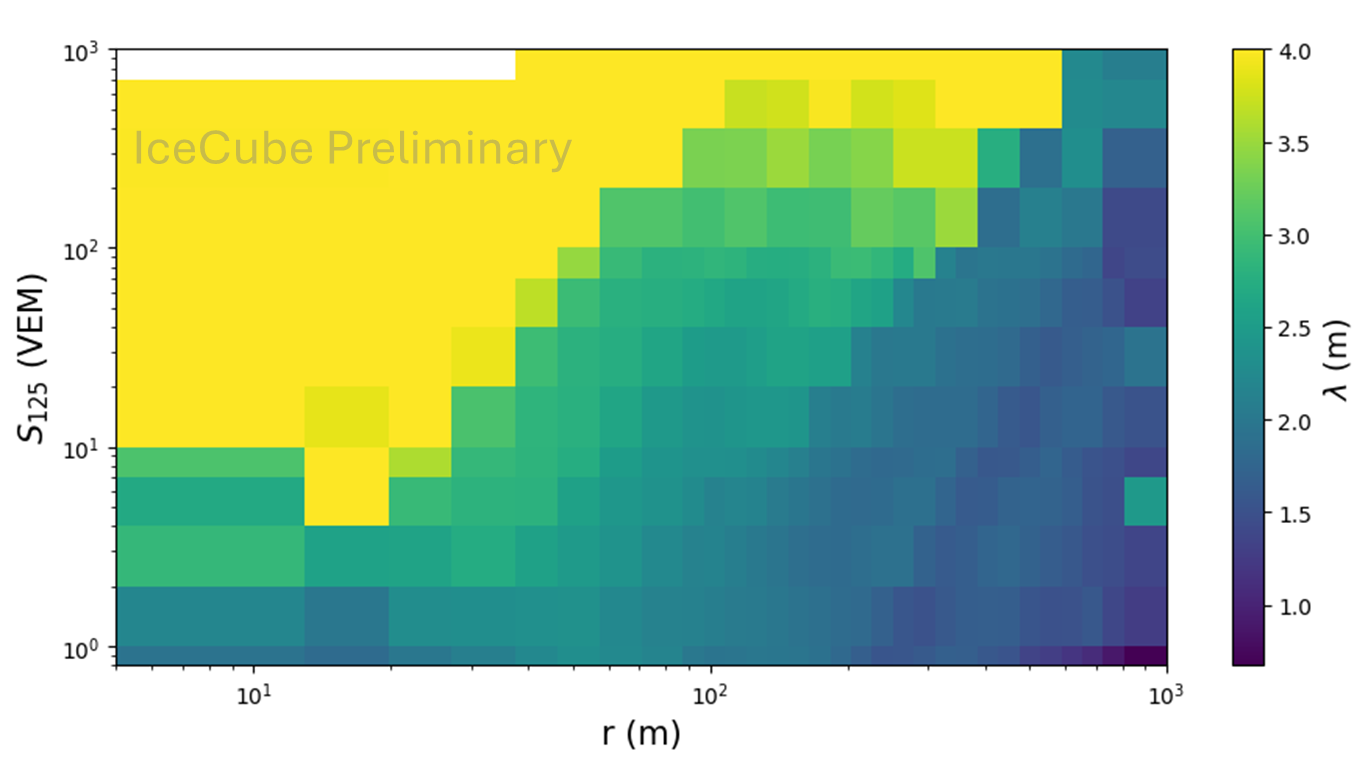}
\caption{$\lambda$ as computed for a set of $r$ and $S_{125}$ ranges. At the highest energies (i.e., for the largest $S_{125}$ and smallest $r$ ranges), statistics are too low to reliably compute a 2-D fit, hence the white space in the upper-left.}\label{fig:snow_lambda_moneyplot}
\end{SCfigure}

Saturation comes into effect at very large charges.
\fixme{NOTE: If we need to cut something from the Snow section, I nominate this paragraph.} For these cases, a modified $\mu_{\text{EM}_\text{sat}}$ is computed (\cref{eq:mu_sat}) as the mean of the saturated distribution, where $S_\text{sat}$ is the value of $S_{\ln}$ above which saturation occurs and $F_\text{under}$ is the fraction of the distribution that falls below $S_\text{sat}$.
\begin{equation}\label{eq:mu_sat}
    \mu_{\text{EM}_{\text{sat}}} = [-\sigma_{\text{EM}}^{2}G(S_{\text{sat}},\mu_{\text{EM}},\sigma_{\text{EM}})+\mu_{\text{EM}}F_{\text{under}}]+[(1-F_{\text{under}})S_{\text{sat}}].
\end{equation}
If the charges are so large that they are saturated at all snow depths, $\lambda$ becomes difficult if not impossible to measure. 
Therefore, in this regime (the upper left of \cref{fig:snow_lambda_moneyplot}), 
$\lambda$ is constrained with an upper limit of 4~meters.

Altogether, each fit has 10 free parameters: 
$\mu_\mu$, $\sigma_\mu$, 
$\mu_{\text{EM}_\text{no snow}}$, $\lambda$, 
$\sigma_{\text{EM}_\text{no snow}}$, $\sigma_{\text{EM}_\text{slope}}$,
$S_\text{thresh}$, $\delta$,
$S_\text{sat}$, and $F_\text{EM}$.
Collecting all of the results from many such fits, we can characterize the behavior of $\lambda$, as well as the fraction of the signal which is from the non-attenuating muon peak, both as a function of \Sonetwentyfive{} and $r$, as is shown in \cref{fig:snow_lambda_moneyplot}.
%
%
Modeling of the muon fraction from these fit results, and testing the overall model on data from multiple years, are both ongoing efforts. It is expected that the resulting reconstruction behavior from early years of IceTop data (when snow coverage was shallow) will match those from later years, across a larger energy range.

\section{Contained Events}
\label{sec:3_CE}
\sectionvspacecheat
The quality cuts used for the last published high-energy spectrum analysis in Ref.~\cite{3year_composition_paper} were mostly based on topological properties, and they aimed to save events that landed inside the borders
of the array. 
In previous work and in this work,
``containment'' of an event is parametrized as the 
distance from the center of the array to the shower core, expressed as a fraction of the distance to the edge of the array in that direction.
In this work, new selection criteria have been 
developed, to enhance energy resolution and increase the event rate, particularly at high energies.

The first of these new cuts is based on the goodness-of-fit from the timing reconstruction: $\chi^2_\mathrm{time}$. The cut value itself depends on the event's total charge, because 
higher-charge events typically involve more triggered tanks, potentially resulting in higher values. 
The second new cut is based on the ratio of
the sum of charges in the outer ring of tanks, to that in the inner array.
This cut is motivated by events with shower cores near the edge of the array, which are only reliably reconstructed if they also deposit substantial charges in the inner tanks.
Like the timing goodness-of-fit cut, the charge-ratio cut also depends on the total charge; misreconstructed events are often characterized by a low total charge, motivating stricter cuts for such events and looser cuts for those with higher total charges. 

\begin{SCfigure}
\includegraphics[width=.5\textwidth]{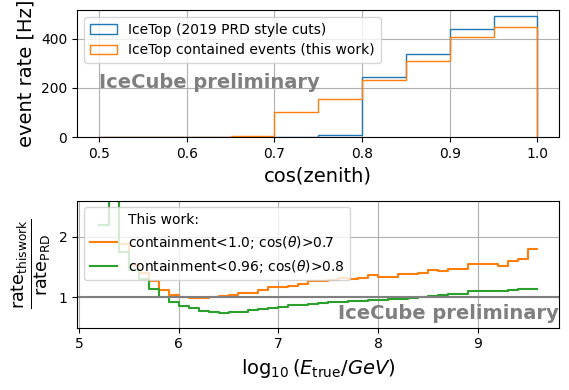}
\caption{An increase of event statistics is achieved by the new selection. \newline
Top: Shows the increase in event rate due to the increased zenith range. Both histograms are weighted to an H4a flux. \newline
Bottom: The ratio of the rates versus the true energy shows an increased rate 
over nearly the whole energy range. When restricting the zenith range and the contained area to the smaller values of the previous analysis for both analyses it is visible that for the highest energies a rate increase over all zenith ranges is achieved.}
\label{fig:CCErate}
\end{SCfigure}

These new cuts allow future analyses to relax some of the selection criteria that were necessary in past analyses. 
For instance, Ref.~\cite{3year_composition_paper} considered
only events from zenith angles of $\cos(\theta) > 0.8$ and containment < 0.96; 
in this work, events out to $\cos(\theta) > 0.7$ and containment < 1.0 are explored.
\begin{SCfigure}[][b]
\includegraphics[width=.5\textwidth]{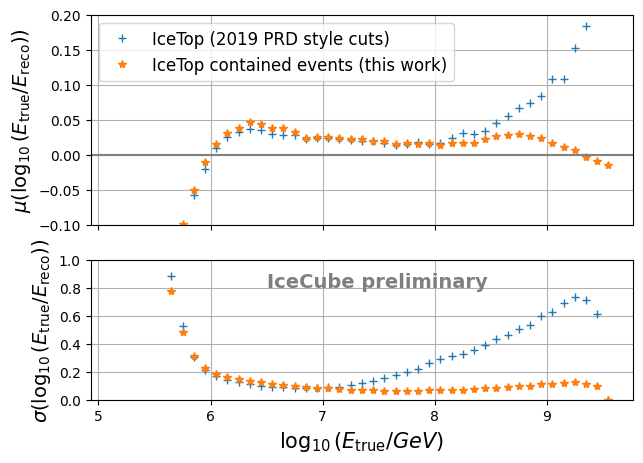}
\caption{Comparison of the energy bias (top) and resolution (bottom) between the previous analysis in Ref.~\cite{3year_composition_paper} and this work. Both show great improvements for the new cuts at the highest energies and similar performance for the lower energies. The slightly increased bias and resolution at the lowest energies are due to the increased zenith range. 
}
\label{fig:CCEresolution}
\end{SCfigure}
A comparison of energy bias and resolution between the previous work and the work presented here is shown in \cref{fig:CCEresolution}. The updated cuts yield significant improvements in resolution at the highest energies, while maintaining similar performance at lower energies. Slightly increased bias and resolution at the lowest energies are attributed to the extended zenith range. 
\cref{fig:CCErate} 
demonstrates a noticeable increase in the event statistics, mainly due to the extension of the zenith range. In the bottom panel of \cref{fig:CCErate}, 
one can see that with the zenith range expansion, an increased rate is expected over the whole energy range. But even when restricting this work to the same zenith range and containment limits as in the previous analysis, 
an increased rate is still expected for the highest energies.
Overall, the newly developed cuts 
provide better energy resolution and higher event statistics. The absolute cut values will be re-optimized when the updated snow model becomes available, which may further improve the results. 
This work presents results from a \textsc{Corsika} simulation dataset which uses the \textsc{Sibyll~2.1} interaction model and the snow overburden corresponding to the year 2012. Comparisons with different interaction models (\textsc{Sibyll~2.3}, \textsc{Sibyll~2.3d}, \textsc{EPOS-LHC} and \textsc{QGSJet-II.04}) and snow overburdens for other years (2015 and 2018) yield similar results.
\section{Uncontained Events}
\label{sec:4_UcE}
\sectionvspacecheat
Previous analyses have required the shower core to land inside the IceTop array, and have been restricted to zenith angles up to 30\degree, which limits the statistics available at the highest energies.
Air showers that land outside the IceTop array are of special interest.
These events typically have no saturated tanks even at high energies, and when combined with the in-ice detector, can explore a larger range of zenith angles.
This can help increase statistics and introduce greater variation in the distance to the shower maximum. 

However, reconstructing uncontained events requires new methods. 
In particular, IceTop-InIce \emph{coincidence} events are analyzed. The in-ice detector records signals from highly energetic muon bundles near the shower axis, providing an anchor for both the shower's core position and direction.
While the electromagnetic component dominates near the shower core and produces HLC hits in IceTop, the signal further from the core is dominated by muons, which typically produce SLC hits. As muons are the primary contributors at large lateral distances, using SLC information becomes crucial for reconstructing air showers with cores far outside the array. To account for this, a \emph{two-component lateral distribution function} (LDF) \cite{2LDF2025, 2LDF2025_Lincoln} is employed that fits the charge distribution for the EM and muonic components separately using HLC and SLC hits. The EM charge distribution is described by the DLP function (\cref{eq:Sem}) with $\kappa=0.30264$ as determined in Ref.~\cite{IceCube:2012nn}. To account for a more stable fit of the EM LDF, a larger reference distance of $r_\text{EM, ref}=400$\,m is chosen, as discussed in \cref{sec:1_intro}. 
The muon contribution is based on the Greisen LDF~\cite{Greisen:1960wc}:
\begin{equation} \label{eq_Smu}
S_{\mu} = S_{\mu,550} \left( \frac{r}{r_{\mu}} \right)^{-\beta_\mu} \left( \frac{r+320\,\rm{m}}{r_{\mu}+320\,\rm{m}} \right)^{-\gamma} , \ r_\mu = 550\,\rm{m}.
\end{equation}
Where $S_{\mu,550}$ is the signal strength at a distance of 550\,m of the shower core and $\beta_\mu$ and $\gamma$ describe the slope of the muon LDF. Details can be found in Ref.~\cite{2LDF2025}.
The reconstruction is performed within the RockBottom framework \cite{RockBottom} and uses a three-step maximum-likelihood method where the charges and timing of the measured signals are compared to the expected charge and timing distributions.
\begin{figure}[t!]
\centering
\includegraphics[width=0.46\linewidth]{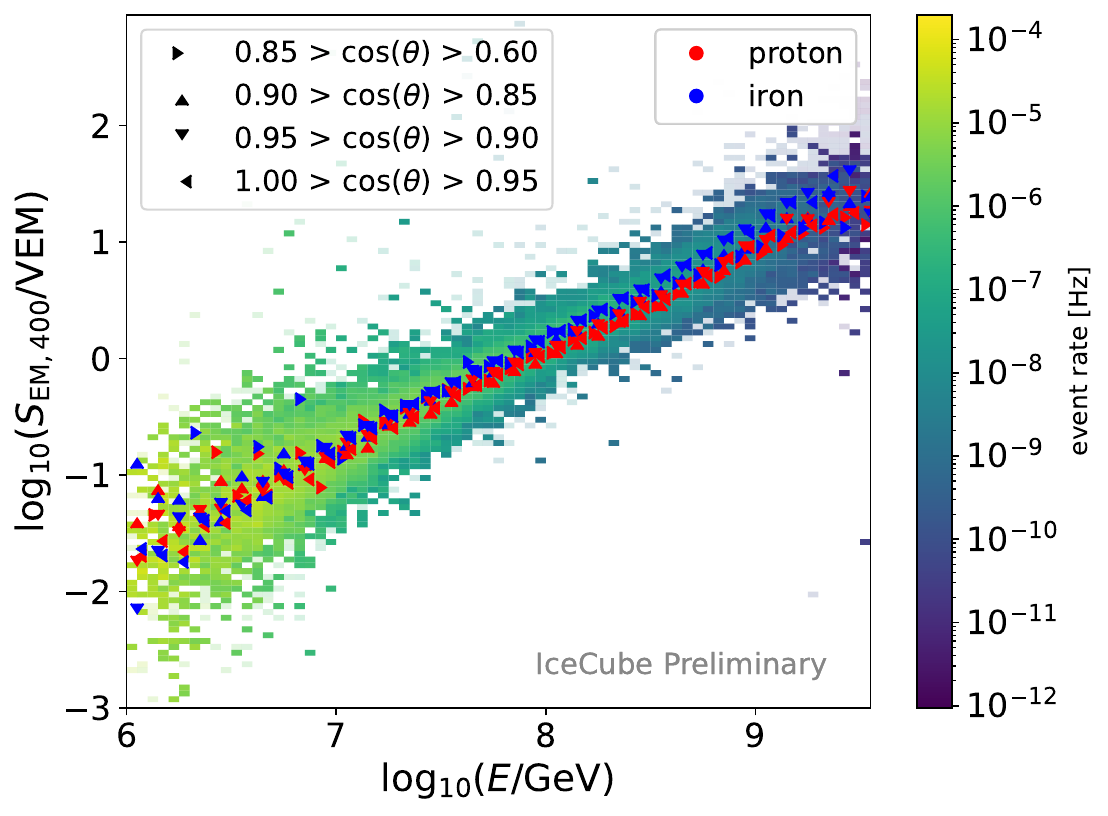}
\includegraphics[width=0.46\linewidth]{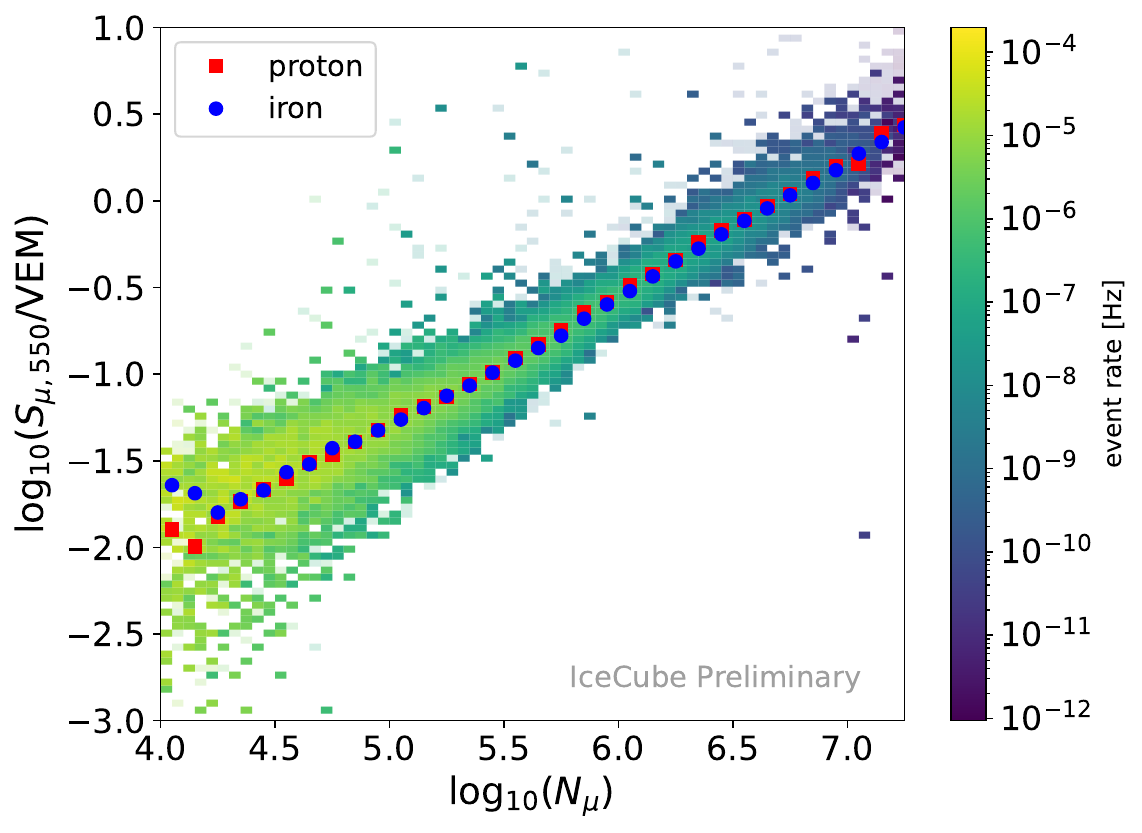}
\caption{Energy proxy, $S_{\rm{EM, 400}}$ (left), and muon number proxy, $S_{\mu,550}$ (right) as a function true energy and true muon number, respectively. The 2d-histogram shows all simulated events (\textsc{Sibyll~2.1}) weighted to a H4a cosmic-ray spectrum. The shaded events are excluded in the analysis. The markers show the mean value for the indicated zenith angles (left) and the complete zenith band (right). The marker color indicates the primary particle.}
\label{fig:uncontained_reco}
\end{figure}
\vspace{-2mm}
\begin{itemize}
\setlength\itemsep{-0em}
\setlength\itemindent{1em}
    \item [\textbf{Step 1:}] 
    An overall fit to a single IceTop LDF ($S_\text{EM}$ and $\beta_\text{EM}$ from 
    \cref{eq:Sem}) as well as the track position and direction ($x,y, \theta, \phi, t$) is performed, using a combined likelihood from both IceTop HLC hits and in-ice hit information.
    $S_\text{EM}$ is seeded with a weighted charge summed over all stations as suggested in Ref.~\cite{Ros:2011zg}.
    \item [\textbf{Step 2:}] With the track position and direction from Step 1 held fixed, a "two-LDF" IceTop reconstruction is performed using both HLC's and SLC's from IceTop alone. The normalization and slope of the EM and muonic LDF~($S_\mathrm{EM}, \beta_\text{EM}$), ($S_\mu, \beta_\mu, \gamma$) are fitted.
    \item [\textbf{Step 3:}] The final step combines the features of Step 1 and Step 2.
    Information from both IceTop (HLC's and SLC's) and the in-ice detector are used for a combined reconstruction of
    a two-LDF model in IceTop and the position and direction of the track.
    All ten parameters---($S_\text{EM}, \beta_\text{EM}$), ($S_\mu, \beta_\mu, \gamma$), and ($x, y, \theta, \phi, t$) ---
    are fitted simultaneously.
\end{itemize}
\vspace{-2mm}
The parameters $S_\text{em}$ and $S_\mu$ are related to the primary cosmic ray energy and the number of muons, respectively. 
Their relationships
are shown in \cref{fig:uncontained_reco}. The simulated events are generated using \textsc{Corsika} \cite{Heck_1998} with the \textsc{Sibyll~2.1} hadronic interaction model \cite{Ahn:2009wx}, assuming the H4a cosmic ray flux model \cite{Gaisser:2011klf}. To ensure reliable in-ice information, only events located outside the 0.9 containment area of IceTop but within the 0.9 containment of IceCube are considered. Furthermore, events must show a coincidence between IceTop and IceCube and exceed a total charge of 100\,PE in both detector components independently.
To enhance the reconstruction quality, events with a total charge below 150\,PE in HLC hits are excluded. Additional selection criteria are applied based on the bounds of the parameters $\beta_\text{EM}$ and $\beta_\mu$, as well as a fit quality requirement. Events not passing these cuts are shown with reduced opacity in \cref{fig:uncontained_reco}. 
The resulting resolution and bias obtained with a cubic fit to the mean values are presented in \cref{fig:uncontained_bias}. In the energy range from $10^7$\,GeV to $10^9$\,GeV, the energy resolution is below~0.2 in $\log_{10} E$. At higher energies, limited statistics reduce reliability. The reconstruction of the muon number shows a better performance in terms of resolution and bias, and is compatible with the results obtained for contained events, as reported in Ref.~\cite{2LDF2025, 2LDF2025_Lincoln}. Ongoing efforts aim to further optimize both bias and resolution. Possible improvements include the use of a zenith- and distance-dependent conversion function, adjustments to the reference distance, and tighter selection cuts. 
These results demonstrate that the reconstruction of uncontained events is feasible. The selection criteria can be tailored to the specific goals of the analysis and offer a valuable complement to the well-established reconstruction of contained events—extending the phase space of reconstructable cosmic rays to larger zenith angles and higher energies.
\begin{figure}[t!]
\centering
\includegraphics[width=0.48\linewidth]{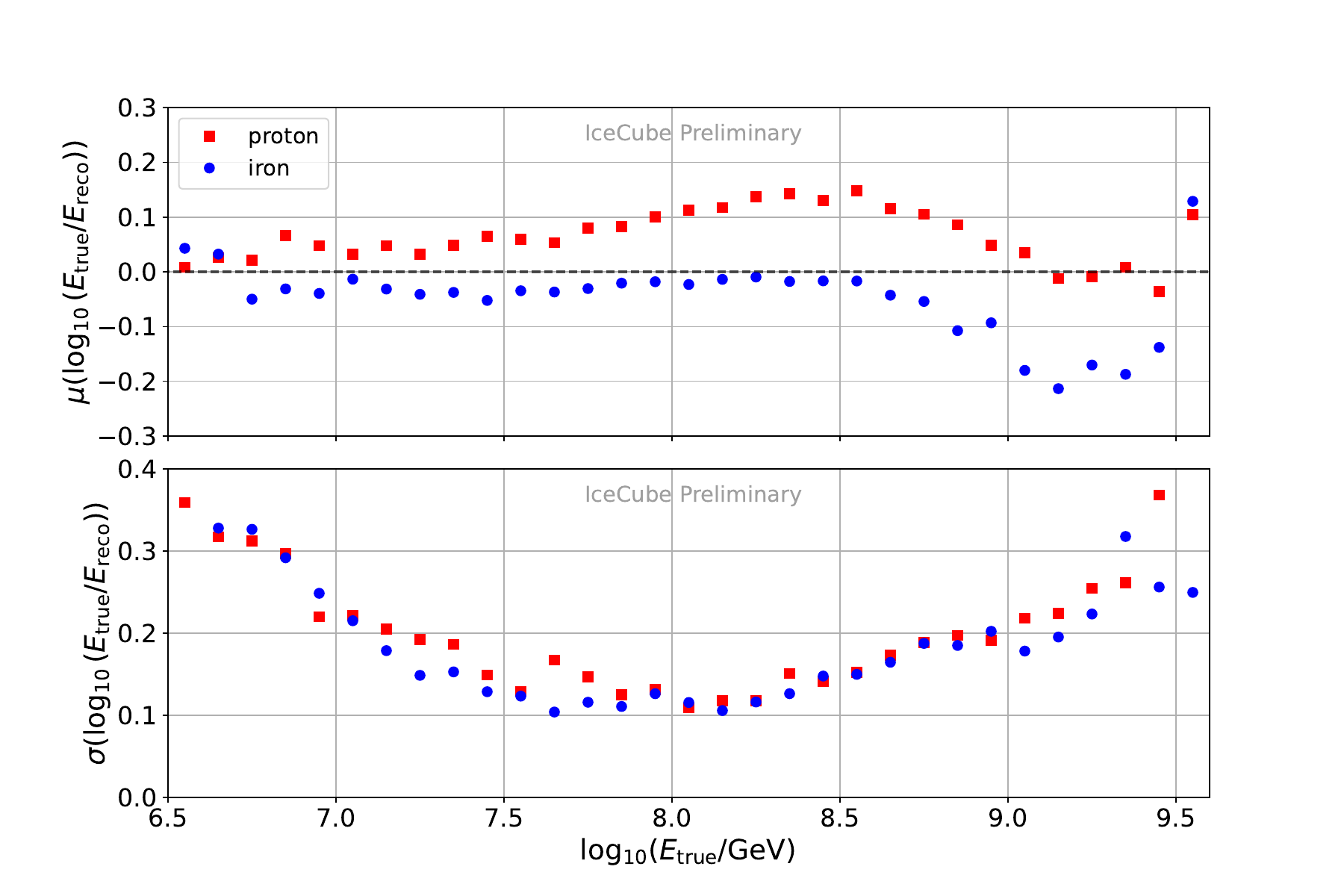}
\includegraphics[width=0.48\linewidth]{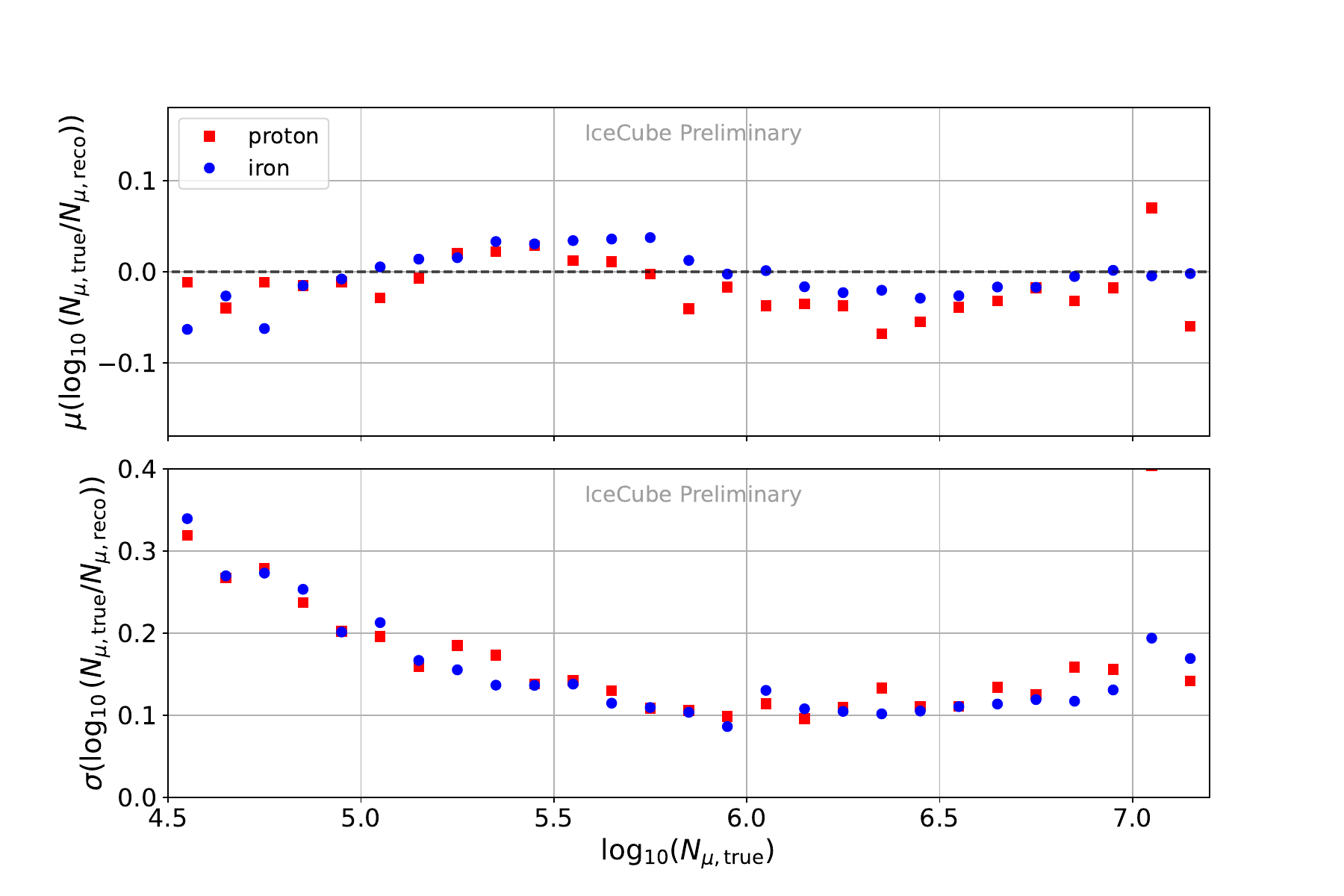}
\caption{Left: Energy bias (upper) and resolution (lower) obtained with a cubic fit to the mean values of the four zenith angular ranges shown in \cref{fig:uncontained_reco}. Right: Muon number bias (upper) and resolution (lower) obtained with one cubic fit.}
\label{fig:uncontained_bias}
\end{figure}
\section{Conclusion}
\sectionvspacecheat
Efforts to enhance IceTop's measurement of the cosmic-ray spectrum are advancing along multiple fronts, aiming to increase statistics at the highest energies and reduce systematic uncertainties in that regime. The newly developed snow model allows for a uniform analysis of more than a decade of data. Improved selection criteria preserve more contained events with higher purity, while the capability to reconstruct uncontained events expands the accessible zenith range, enabling future analyses of the energy spectrum and composition of high-energy cosmic rays.

\bibliographystyle{ICRC}
\bibliography{references}
\clearpage
\section*{Full Author List: IceCube Collaboration}

\scriptsize
\noindent
R. Abbasi$^{16}$,
M. Ackermann$^{63}$,
J. Adams$^{17}$,
S. K. Agarwalla$^{39,\: {\rm a}}$,
J. A. Aguilar$^{10}$,
M. Ahlers$^{21}$,
J.M. Alameddine$^{22}$,
S. Ali$^{35}$,
N. M. Amin$^{43}$,
K. Andeen$^{41}$
% ...
\\
\\
% ...
$^{10}$ Universit{\'e} Libre de Bruxelles, Science Faculty CP230, B-1050 Brussels, Belgium \\
% ...
$^{16}$ Department of Physics, Loyola University Chicago, Chicago, IL 60660, USA \\
$^{17}$ Dept. of Physics and Astronomy, University of Canterbury, Private Bag 4800, Christchurch, New Zealand \\
% ...
$^{21}$ Niels Bohr Institute, University of Copenhagen, DK-2100 Copenhagen, Denmark \\
$^{22}$ Dept. of Physics, TU Dortmund University, D-44221 Dortmund, Germany \\
% ...
$^{35}$ Dept. of Physics and Astronomy, University of Kansas, Lawrence, KS 66045, USA \\
% ...
$^{39}$ Dept. of Physics and Wisconsin IceCube Particle Astrophysics Center, University of Wisconsin{\textemdash}Madison, Madison, WI 53706, USA \\
% ...
$^{41}$ Department of Physics, Marquette University, Milwaukee, WI 53201, USA \\
% ...
$^{43}$ Bartol Research Institute and Dept. of Physics and Astronomy, University of Delaware, Newark, DE 19716, USA \\
% ...
$^{63}$ Deutsches Elektronen-Synchrotron DESY, Platanenallee 6, D-15738 Zeuthen, Germany \\
$^{\rm a}$ also at Institute of Physics, Sachivalaya Marg, Sainik School Post, Bhubaneswar 751005, India \\
% ...

\subsection*{Acknowledgments}

\noindent
The authors gratefully acknowledge the support from the following agencies and institutions:
USA {\textendash} U.S. National Science Foundation-Office of Polar Programs,
U.S. National Science Foundation-Physics Division,
U.S. National Science Foundation-EPSCoR,
U.S. National Science Foundation-Office of Advanced Cyberinfrastructure,
Wisconsin Alumni Research Foundation,
Center for High Throughput Computing (CHTC) at the University of Wisconsin{\textendash}Madison,
Open Science Grid (OSG),
Partnership to Advance Throughput Computing (PATh),
Advanced Cyberinfrastructure Coordination Ecosystem: Services {\&} Support (ACCESS),
Frontera and Ranch computing project at the Texas Advanced Computing Center,
U.S. Department of Energy-National Energy Research Scientific Computing Center,
Particle astrophysics research computing center at the University of Maryland,
Institute for Cyber-Enabled Research at Michigan State University,
Astroparticle physics computational facility at Marquette University,
NVIDIA Corporation,
and Google Cloud Platform;
Belgium {\textendash} Funds for Scientific Research (FRS-FNRS and FWO),
FWO Odysseus and Big Science programmes,
and Belgian Federal Science Policy Office (Belspo);
Germany {\textendash} Bundesministerium f{\"u}r Forschung, Technologie und Raumfahrt (BMFTR),
Deutsche Forschungsgemeinschaft (DFG),
Helmholtz Alliance for Astroparticle Physics (HAP),
Initiative and Networking Fund of the Helmholtz Association,
Deutsches Elektronen Synchrotron (DESY),
and High Performance Computing cluster of the RWTH Aachen;
Sweden {\textendash} Swedish Research Council,
Swedish Polar Research Secretariat,
Swedish National Infrastructure for Computing (SNIC),
and Knut and Alice Wallenberg Foundation;
European Union {\textendash} EGI Advanced Computing for research;
Australia {\textendash} Australian Research Council;
Canada {\textendash} Natural Sciences and Engineering Research Council of Canada,
Calcul Qu{\'e}bec, Compute Ontario, Canada Foundation for Innovation, WestGrid, and Digital Research Alliance of Canada;
Denmark {\textendash} Villum Fonden, Carlsberg Foundation, and European Commission;
New Zealand {\textendash} Marsden Fund;
Japan {\textendash} Japan Society for Promotion of Science (JSPS)
and Institute for Global Prominent Research (IGPR) of Chiba University;
Korea {\textendash} National Research Foundation of Korea (NRF);
Switzerland {\textendash} Swiss National Science Foundation (SNSF).
\end{document}